# Challenges in Band Alignment between Semiconducting Materials: A Case of Rutile and Anatase TiO$_2$

Daoyu Zhang[*] and Shuai Dong[†]

School of Physics, Southeast University, Nanjing, 211189, China

**ABSTRACT**

This topical review focuses on the recently active debate on the band alignment between two polymorphs of TiO$_2$, rutile and anatase. A summary is given to the popular methods for measurement and calculation of band alignment between materials. We point out, through examination of recently experimental and theoretical reports, that the outstanding discrepancy in the band alignment between two TiO$_2$ phases is attributed to factors that influence band alignment rather than needs a definite answer of which band alignment is right. According to an important factor, the presence of an interface, a new classification of band alignment is proposed as the coupled and intrinsic band alignments. This classification indeed reveals that the rutile/anatase interface can qualitatively change the type of their band alignment. However, further systematic information of the interface and other factors that influence band alignment will be needed to understand changes in energy bands of materials better. The results obtained from discussion of the band alignment between rutile and anatase may also work for the band alignment between other semiconductors.

[*] Email: zhangdaoyu@seu.edu.cn

[†] Email: sdong@seu.edu.cn

## 1. Introduction

Utilizing solar irradiation to degrade water into hydrogen and oxygen gases is a well-established idea to produce green energy [1-6]. This degradation process originates from the band gap of a semiconductor used in photocatalysis. Photons with higher energy than the semiconductor band gap can be absorbed by the semiconductor, and simultaneously electrons in the valence band are excited to the conduction band and holes are left in the valence band. These electron-hole pairs are chemically active, and therefore they can be directly used to initiate chemical redox reactions. The utilizing efficiency of solar irradiation depends on the band gap of a photocatalytic material. Besides the band-gap requirement during the photocatalytic process, there is the band-edges requirement.

There is a thermodynamic relationship of the band edges of a semiconductor photocatalyst and the redox potentials of species entering into redox reactions. Fig. 1 illustrates the alignment between band edges of the anatase $TiO_2$ and the electrochemical potentials of the two redox half reactions of water decomposition with respect to the vacuum level [7]. It is clear while driving the oxygen evolution reaction implies the photo-generated-holes energy must be lower than the electrochemical potential of the $O_2/H_2O$ redox couple, and driving the hydrogen evolution reaction requires the photogenerated-electrons energy must higher than the $H^+/H_2$ redox couple. The thermodynamic rule states that the higher the conduction band minimum energy and the lower valence band maximum energy, the stronger tendency for reduction and oxidation reactions respectively. The ideal energy difference between the valence and conduction band edges for photoelectrolysis of water is frequently reported as 1.6-2.4 eV, when taking the losses at the semiconductor/liquid interface due to the concentration and kinetic overpotentials needed to drive reactions into consideration, and the two band edges must straddle the electrochemical potentials of the $O_2/H_2O$ and $H^+/H_2$ redox couples [8].

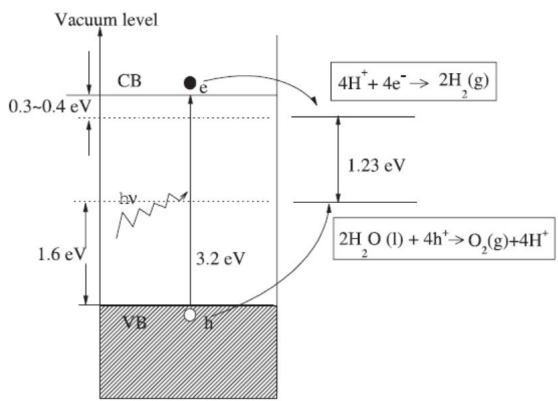

**Fig. 1.**The photocatalytic reducing and oxidation powers of a semiconductor (e.g. anatase $TiO_2$) depend on the conduction band minimum and the valence band maximum, respectively. Reprinted figure with permission from Gai*et al.*[7].Copyright © (2009) by the American Physical

Society.

Generally, the photo-generated electron-hole recombination can reduce the efficiency of the decomposition of water or other compounds [9]. Numerous works have devoted their efforts to effectively separate electron-hole pairs [10, 11]. Hetero-structures are good choices for electron-hole separation and are benefit to the transfer of charge carriers [12, 13], because the internal electric field develops in the depletion layer. Electron-hole pairs in the depletion layer, no matter created by light in the depletion layer or diffused from other part, can be spatially separated by the internal electrical field. Subsequently, electrons and holes drift to the opposite sides of the hetero-junction, respectively. Therefore, electrons and holes have long lifetimes enough to reach the catalyst's surfaces and to drive the redox reactions.

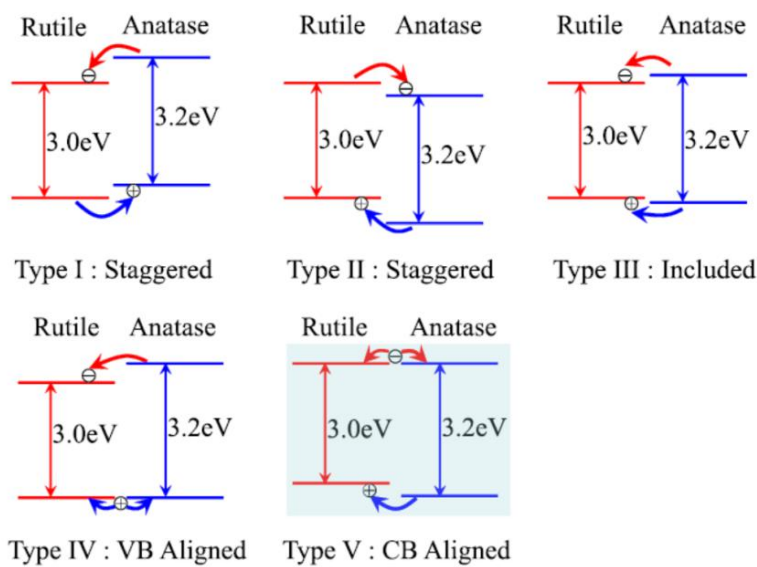

**Fig. 2.** Schematic five possible types of the band alignment between rutile and anatase TiO$_2$. Reprinted by permission from Macmillan Publishers: Mi *et al.*[14]. Copyright © (2015).

TiO$_2$ has two major stable phases: rutile and anatase. The hetero-structure constructed by rutile and anatase TiO$_2$ shows higher photocatalytic activity than their pristine ones. It has been generally accepted that the photo-generated electrons and holes move to opposite sides of the junction. However, the transfer directions of the charge carriers remain under debated: one opinion is that the electrons move from anatase to rutile, while the holes from rutile to anatase; another viewpoint is totally opposite. From the band alignment of view, there are five possible band alignments between rutile and anatase, as sketched in Fig. 2, where the transfer directions of the charge carriers are marked by arrows [14]. The type I denotes the staggered alignment with anatase band edges lying above rutile ones [15, 16]; the type II denotes the staggered alignment with rutile band edges lying above anatase ones [17, 18]; the type III is the straddled alignment [19-21]; the type IV has the same position of the valence band edge [22-24]; and the type V has the same position of the conduction band edge [25, 26]. Each of five types has its own theoretical and experimental support, making the band alignment of rutile and anatase $TiO_2$ be an open question under much debate.

## 2. Experimental measurements of band alignment

There are two popular measuring methods for the band alignment between two materials, electrochemical impedance analysis (EIA) and X-ray photoelectron spectroscopy (XPS). Both methods have been employed to characterize the band alignment between rutile and anatase $TiO_2$, which will be acted as a demo of methods.

*2.1. Electrochemical impedance analysis (EIA) measurement*

Since $TiO_2$ is an *n*-type semiconductor, the following EIA measurement theory is based on the *n*-type semiconductor, and similar situation works for *p*-type semiconductor. Fig. 3(a) shows the electronic structures of an *n*-type semiconductor and the electrolyte solution containing redox species before their contact with each other. When the semiconductor is in contact with the electrolyte solution, electrons will transfer across the semiconductor/solution interface until the Fermi levels of two phases are equal, as shown in Fig. 3(b). Given that the Fermi level of the semiconductor ($E_F$) is higher than that of the redox couple ($E_{F(redox)}$), the transfer of electrons from the semiconductor to the solution creates a positive space charge layer in the semiconductor side, which bends the energy band of the semiconductor. Furthermore, the Helmholtz double layer in the solution side develops simultaneously, which leads to an additional potential drop ($V_H$). $V_H$ is a component of the measured electrode potential of the semiconductor. When an external voltage on the semiconductor is applied, the band-bending changes with the applied voltage. If the applied voltage is adjusted just right to getting the band of the semiconductor back to no band bending (see Fig. 3(a)), the measured electrode potential of the semiconductor is so-called the flat band potential ($U_{fb}$), a critical parameter for band alignment using EIA.

The flat band potential can be determine by Mott-Schottky equation [27]:

$$\frac{1}{C^2} = \frac{2}{\varepsilon\varepsilon_0 A^2 eN_D}\left(V - U_{fb} - \frac{k_b T}{e}\right), \tag{1}$$

where $C$ is the interfacial capacitance; $\varepsilon$ is the dielectric constant of the semiconductor; $\varepsilon_0$ is the permittivity of free space; $A$ is the interfacial area; $N_D$ is the number of donors; $V$ is the applied voltage; $k_b$ is the Boltzmann's constant; $T$ is the absolute temperature; and $e$ is the electronic charge. A plot of $1/C^2$ against $V$ should be a straight line, whose intercept on the $V$ axis can give the flat potential.

The flat potential $U_{fb}$ links the energy levels of the semiconductor and electrolyte by the relationship[4, 28]:

$$U_{fb} = \chi + \Delta E_F + V_H + \Phi_0, \quad (2)$$

where $\chi$ is the electron affinity of the semiconductor; $\Delta E_F$ is the difference between the Fermi level and majority carrier band edge of the semiconductor; and $\Phi_0$ is the scale factor (e.g. 4.5 eV for NHE, the normal hydrogen electrode). When $V_H$ equals zero at zero point of charge (pH$_{ZPC}$), the flat band potential (denoted by $U_{fb}^0$) is the intrinsic Fermi level of the semiconductor. Only $U_{fb}^0$ is the meaningful flat potential for band alignments mentioned here.

In metal oxides, such as TiO$_2$, the flat band potential varies with the pH value of solution following a linear relation known as the Nernstian relation[28]:

$$U_{fb}^{pH} = U_{fb}^0 + 0.059(pH_{ZPC} - pH). \quad (3)$$

Using measured data of $U_{fb}^{pH}$ and $E_G$ ($E_G$ is the band gap), the valence band edge energy of an $n$-type semiconductor with respect to the reference electrode can be calculated as:

$$E_{VBE} = U_{fb}^{pH} - \Delta E_F + E_G - 0.059(pH_{ZPC} - pH). \quad (4)$$

To determine the value of $\Delta E_F$ experimentally, certain impurities will be introduced into the semiconductor, which can shift the Fermi level so close to the majority carrier band edge that $\Delta E_F$ is approximately zero and may be neglected in Eq. 4.

Using Eq. 4, the band alignment between rutile and anatase TiO$_2$ can be determined. The flat band potential of anatase TiO$_2$ (101) surface relative to the saturated calomel electrode (SCE) is -0.4 eV at pH = 0[24], while that of rutile TiO$_2$ (110) surface is -0.25 eV at pH = 1[29]. Considering the band gaps of 3.2 eV for anatase and of 3.0 eV for rutile, the valence band edge energy of rutile is 0.04 eV lower than that of anatase. The EIA measurement obtains the type IV band alignment for two phases of TiO$_2$.

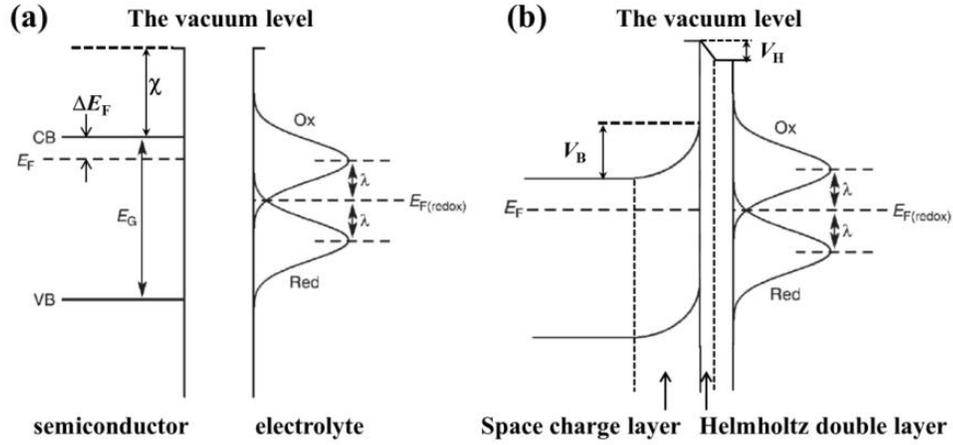

**Fig. 3.** Schematics of energy levels of an *n*-type semiconductor and an electrolyte solution before (a) and after (b) contact with each other, showing the relationships of the Fermi level of the semiconductor ($E_F$), the Fermi level of the redox couple ($E_{F(redox)}$), the electron affinity of the semiconductor ($\chi$), the band bending ($V_B$), the Helmholtz layer potential drop ($V_H$), and the reorganization energy ($\lambda$).

*2.2. X-ray photoelectron spectroscopy (XPS) measurement*

Kraut *et al*. proposed a XPS method to precisely determine the band alignment between any materials [30]. The schematic diagram of this method is shown in Fig. 4. When a semiconductor is in contact with the other material, the space charge layer develops with a typical width of ~100 nm (Fig. 4(a)), as a result, the band bending occurs at the interface. The electronic states with the flat bands at the interface are shown in Fig. 4(b). Note that in the space charge layer all bands or energy levels will be bent by the same value as a function of distance away from the interface according to Poisson's equation. It follows from Fig. 4(a) and 4(b) that the valence band offset of two semiconductors can be given by:

$$\Delta E_v = \left(E_{CL}^Y - E_v^Y\right) - \left(E_{CL}^X - E_v^X\right) - \Delta E_{CL}. \tag{5}$$

Here $\left(E_{CL}^Y - E_v^Y\right)$ and $\left(E_{CL}^X - E_v^X\right)$ are the binding energy difference between the core level and the valence band maximum, being gained by an XPS measurement respectively; and $\Delta E_{CL} = E_{CL}^Y(i) - E_{CL}^X(i)$, which can be determined by an XPS measurement of a hetero-junction composed by materials X and Y.

The band alignment between rutile and anatase TiO$_2$ reported by Scanlon *et al.* will be taken an example for the XPS measurement processing, [18]. The Ti's 2$p_{3/2}$ is taken as the core level. First, $E_{CL}^{rutile} - E_v^{rutile}$ and $E_{CL}^{anatase} - E_v^{anatase}$ are measured from the phase-pure rutile and anatase samples (see Fig. 5(a) and 5(b) respectively. In Fig. 5(a), the energy position of valence band maximum is determined by extrapolating the linear portion of the low energy edge of the valence band to the spectral baseline, which gives the valence band maximums 2.77 eV and 2.61 eV relative to the Fermi level for anatase and rutile respectively. Second, to obtain the value of $\Delta E_{CL} = E_{CL}^{rutile} - E_{CL}^{anatase}$, three high quality rutile-anatase junctions are prepared with the rutile to anatase ratios of 2:1, 1:1 and 1:2 respectively. The spectra (see Fig. 5(c)) from junctions contain the contributions from rutile Ti's 2$p_{3/2}$ and anatase Ti's 2$p_{3/2}$ states, which can be separated by a fit of the Gaussian-Lorentzian function. The average value of $\Delta E_{CL}$ over three junction samples is -0.44 eV. Fig. 5(d) shows the band alignment between rutile and anatase resulting from the XPS measurement. The rutile valence band with respect to the vacuum level was found to be a higher energy position of 0.39 eV than that of the anatase. Therefore, the XPS measurements obtain the type II band alignment for these two phases of TiO$_2$.

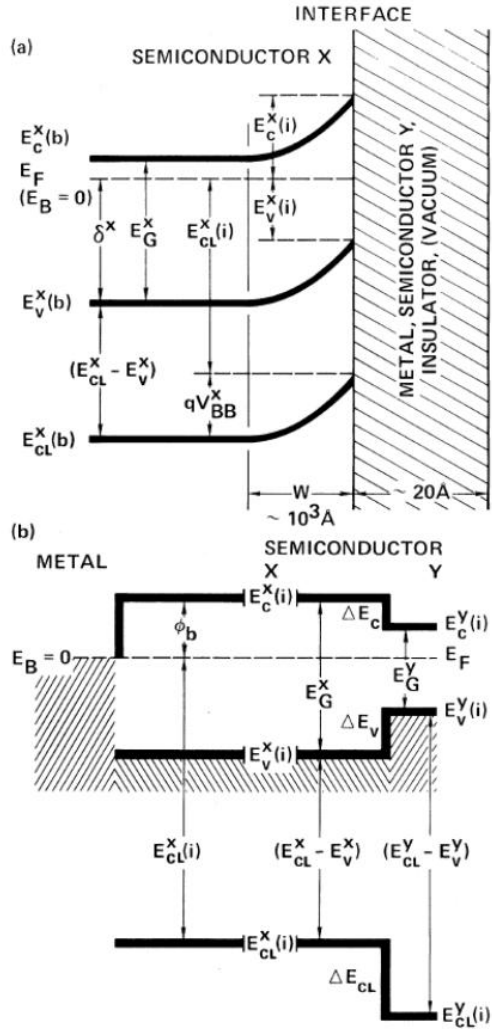

**Fig. 4.**(a) Generalized energy band diagram at a sharp interface between a semiconductor and vacuum, metal, insulator, or a different semiconductor. Binding energy $E_B$ is measured with respect to the Fermi level $E_F$. (b) Schematic flat band diagram at metal-semiconductor (left) or hetero-junction (right) interface. Reprinted figure with permission from Ref.[30]. Copyright © (1980) by the American Physical Society.

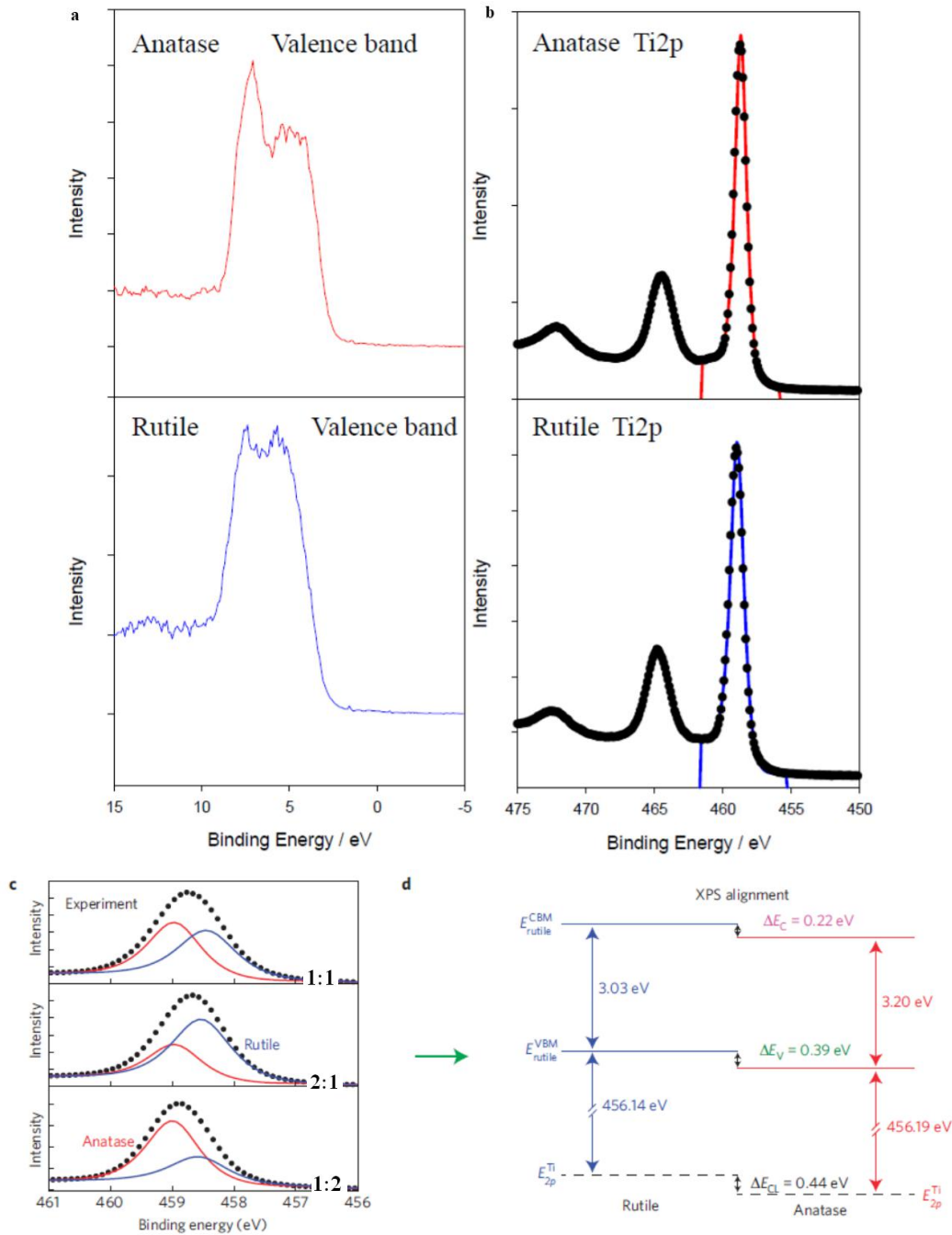

**Fig. 5.** Valence band (a) and Ti's $2p_{3/2}$ (b) XPS spectra measured from pure phase anatase and rutile samples. The core level spectra are fitted by a single Gaussian-Lorentzian function. (c) Ti's $2p_{3/2}$ spectra measured from the junctions with rutile to anatase ratios of 1:1, 2:1 and 3:1. The two curves are the fitted curves of Gaussian-Lorentzian functions. (d) Schematic band alignment between rutile and anatase $TiO_2$ from the XPS measurement. Reprinted by permission from Macmillan Publishers Ltd: Scanlon *et al.* [18]. Copyright © (2013).

## 3. Theoretical calculations of material band alignments

Any theoretical approach to determine the band alignment between two materials faces a fundamental issue: choosing 'a certain electronic state' as the common reference level to align energy bands of two materials. Such 'an electronic state' may be a localized core level, the vacuum level, the averaged electrostatic potential, the charge neutrality level, the common anion rule, atomic electronegativity, or possible others [17]. The latter three alignment references are not exactly accurate, and they predict band alignments only for part materials being consistent with experimental data [31]. For example, the method of atomic electronegativity calculates the band alignments between materials only using their chemical composition [32, 33]. In this way, polymorphs of a solid material, such as $TiO_2$, rutile and anatase, should have the same positions of the energy band edges, so that the results are not consistent with experiment ones. Therefore, three approaches for alignment will be introduced, which can be implemented by first-principles calculations.

*3.1. The method based on the core-level reference*

Similar to the XPS measurement procedure for band alignments between materials, three steps can determine the valence band alignment between materials A and B. First, the energy spacing between the valence band maximum and the core level ($\Delta E_{v,C}$ shown in Fig. 6), is calculated at its respective equilibrium lattice constants. Second, the energy difference between the core levels of A and B, $\Delta E_{C,C'}(A/B)$, is obtained from a superlattice composed by A and B. Note that the superlattice needs to be unrelaxed to prevent the shifts of core levels. Third, the valence band offset, $\Delta E_{v,C}$, can be derived using Eq. 5.

This method has been widely used in band alignment calculations with several derivatives. A major difference between derivatives is the ways to construct superlattices. One way is to use the quantum dots (QDs) to create the superlattice, which contains a pair of QDs, i.e. one rutile $TiO_2$ QD and one anatase $TiO_2$ QD, separated by a sufficient vacuum spacing, as shown in Fig. 7(a) [34]. The surface dangling bonds of QDs should be passivated by some atoms (e.g. pseudohydrogen atoms with a fractional nuclear charge), which must be relaxed to their proper positions. In addition, $\Delta E_{C,C'}(A/B)$ determined by a pair of QDs varies with the sizes of the QDs, so the QD size should be carefully tested to converge $\Delta E_{C,C'}(A/B)$ to a constant. Another way to construct the superlattice is using a hetero-structure with the average lattice constants of A and B, as shown in Fig. 7(b) [35]. The thickness of the slab A and B needs to be checked but generally

~1 nm is sufficient to converge $\Delta E_{C,C'}(A/B)$ to within 0.01 eV. Note that the hetero-structure is created by two mismatched lattices, the core levels in the hetero-structure should be corrected. This correction of the core level can be obtain by calculating a series of absolute uniaxial deformation potentials for the core level and performing an angular average [36], or by directly comparing the core level of the original A (or B) with that of A (or B) in the hetero-structure [37]. In addition, the hetero-structure does not mind whether there is the vacuum spacing or not to separate A and B.

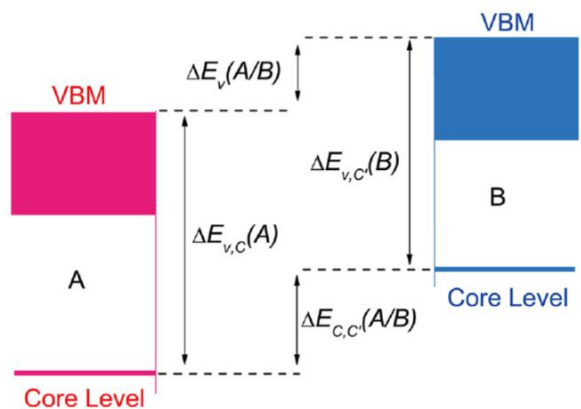

**Fig. 6.** Illustration of the scheme to calculate the band offset using the core level reference. Reprinted with permission from Kang *et al.*[34]. Copyright © (2008) by the American Chemical Society.

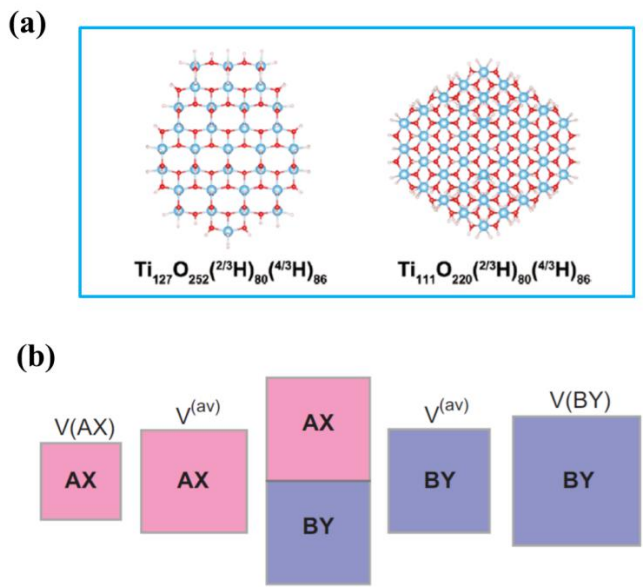

**Fig. 7.** Schematic of a pair of rutile and anatase TiO$_2$ QDs (a) and a rutile/anatase hetero-structure

(b) used to calculate their core level difference. (a) Reprinted with permission from Kang *et al.*[34]. Copyright © (2012) by the American Chemical Society. (b) Reprinted with permission from Li *et al.*[35]. Copyright © (2009) by the American Institute of Physics.

*3.2. The method on basis of the averaged electrostatic potential*

The electrostatic potential here means the Hartree potential, because the ionic potential is implicitly accounted for in the bulk band edge difference. The electrostatic potential at the point $\vec{r}$ is determined by solving the Poisson equation with the electron density from density functional theory (DFT) calculation. The plane-averaged potential $\bar{V}(z)$ across the interface is defined by an integral [38, 39]:

$$\bar{V}(z) = \frac{1}{S}\int_S V(\vec{r})\,dxdy, \tag{6}$$

where S is the area of the plane parallel to *xy* plane in a unit cell. To smooth the oscillation of $\bar{V}(z)$, the macroscopic average $\bar{\bar{V}}(z)$ is defined as:

$$\bar{\bar{V}}(z) = \frac{1}{L}\int_{z-L/2}^{z+L/2} \bar{V}(z')\,dz', \tag{7}$$

where $L$ is the oscillation period, and thus $\bar{\bar{V}}(z)$ tends to be a constant. Based on the macroscopic average of the electrostatic potential, the valence band offset $\Delta E_v$ relative to the vacuum level can be obtained by:

$$\Delta E_v = \left(e\bar{\bar{V}}^Y - E_v^Y\right) - \left(e\bar{\bar{V}}^X - E_v^X\right) + e\Delta\bar{\bar{V}}^{Y/X}, \tag{8}$$

where $\Delta\bar{\bar{V}}^{Y/X} = \bar{\bar{V}}^Y - \bar{\bar{V}}^X$ obtained in the hetero-junction composed by X and Y.

*3.3. The method on basis of the vacuum level*

When a material is in low-dimensional, the vacuum level may be conveniently acted as the common zero energy reference and evaluated by the mean electrostatic potential [40, 41]. For example, the vacuum level for a two-dimensional rutile $TiO_2$ (110) surface is obtained as follows. First, a DFT static calculation gives the electrostatic potential distribution of the surface. Second, the (100)-planar average electrostatic potential varies with the z direction perpendicular to the surface. The average electrostatic potential in the vacuum far away from the surface tends to be a constant (denoted by $E_{vac}$). $E_{vac}$ is the vacuum level. Finally, the absolute position of the valence band edge is given by

$$E_v = E_{VBM} - E_{vac}, \tag{9}$$

where $E_{VBM}$ is the valence band maximum taking from the interior atoms of the surface slab. According to Eq. 9, the valence band offset of the two slabs can obtained simply by subtracting $E_v$ from one another. Therefore, the procedure is much concise when the vacuum level as the common energy reference without a need of creating hetero-strucuture.

## 4. Explanation for discrepancy in the band alignment between rutile and anatase

Theoretical calculations and experimental observations show the different band alignment of rutile and anatase TiO$_2$. How to clarify this apparent discrepancy? There are several viewpoints on this issue.

*4.1. Effect of the dipole layer created by surface adsorption*

As mentioned above, the flat band potential varies with pH following a linear relation, i.e., the Nernstian relation (see Eq. 3). It is widely accepted that this relation results from adsorption of H$^+$ or/and OH$^-$ on the material surface close to the Helmholtz layer. However, the Nernstian relation does not make any distinction between rutile and anatase TiO$_2$. Kullgren*et al.*'s work on adsorption of H$^+$, OH$^-$, and even the water at the rutile (110) and anatase (101) surfaces shows that the adsorption effect of the same species on the band edges is different for rutile and anatase[42]. Adsorption of water only changes the valence band edge significantly for two surfaces but with a similar magnitude, as a result, it does not affect the band alignment of two phases. H$^+$ adsorbed on O$_{2c}$ (two-coordinate O atom) remarkably affects the conduction band alignment. Furthermore, the CBM offset between two phases can be tuned by the number of OH$^-$ groups adsorbed on Ti$_{5c}$ (five-coordinate Ti atom), and the tuned amount falls within the range of the EIA observations. Hence, Kullgren *et al.* conclude that the main difference between XPS and EIA results comes from that the former is carried out in vacuum and the latter in anaqueous electrolyte solution with adsorbing OH$^-$ and H$^+$ ions.

TiO$_2$ materials always have defects depending on the preparation conditions and the post-treatment processes. One of our work indicates that the defects on the rutile (110) and anatase (101) surfaces can change the type of the band alignment between rutile and anatase relative to pure surfaces [43]. Three surface defects, oxygen vacancies (O-vacs) formed by losing bridging oxygen atoms, hydroxyl groups (O-Hs), and surface fluorination (Ti$_{5c}$-Fs) have been considered. When two surfaces are free of defects, the band alignment between rutile and anatase is the type-IV, as shown in Fig. 2. The reductant defects, O-Hs and O-vacs, turn the type of the band alignment to the type II, whereas the oxidative defects, Ti$_{5c}$-Fs[6], lead the band alignment to the type I. The surface electric dipole layer induced by defects is responsible for the type reversal of the band alignment between rutile and anatase TiO$_2$ [43-45].

*4.2. The direct or indirect band gap entering the band alignment*

The rutile TiO$_2$ is a direct band gap semiconductor of 3.0 eV, whereas the anatase TiO$_2$ is an indirect band gap semiconductor [17]. Nosaka *et al.* proposed a mechanism to settle disputes on the band alignment between rutile and anatase from experimental observations [46], as sketched in Fig. 8. The type-II band alignment (in Fig. 2) is due to the effect from the indirect band gap of 3.2 eV, whereas the type-IV alignment comes from the direct band gap of 3.8 eV. Why do those electronic states below the direct conduction band minimum of the anatase phase not enter the EIA band alignment between rutile and anatase? Nosaka *et al.* argued that the small change corresponding to the indirect band probably was overlooked in the measurement of the Mott-Schottky plot reported by Kavan *et el.* [24], when the potential was scanned from negative to positive.

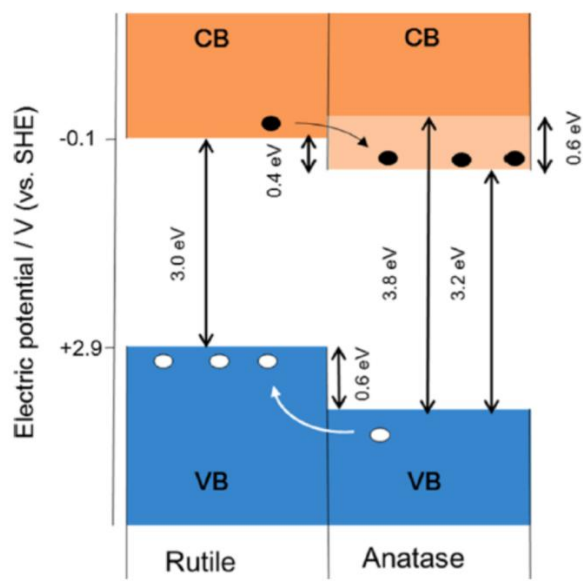

**Fig. 8.** Schematic of the direct or indirect band gap of anatase enters the band alignment between rutile and anatase TiO$_2$. Reprinted with permission from Nosaka *et al.*[46]. Copyright © (2016) by the American Chemical Society.

*4.3. Effect of an interface*

A reliable *ab initio* calculation method for the band alignment between solids A and B has been proposed by us, as illustrated in Fig. 9(a), in which the corrections of the lattice mismatch and the surface polarity are taken into consideration [37]. Similar to the XPS measurement procedure, $\Delta E_{VBM-CL}(A)$ and $\Delta E_{VBM-CL}(B)$ is calculated at their respective equilibrium lattice constants, while $\Delta E_{B-A}(CL)$ –from the A'/B' junction with corrections 1 and 2, which are from the surface polarity and the expansion and/or compression of A and B, respectively. The surface polarity relative to the bulk one is described by the electric dipole moment, $p$, and the moment of the dipole layer induced by the interaction between two solids is denoted by $p_{inter}$. This method can reveal the effect of the interaction between solids on their band alignment. When the separation spacing, $d$, between two surface slabs is more than 5 Å, the band alignment between rutile and anatase is independent on $d$. In contrast, the distances $d$ less than 5 Å lead to a notable change in the band alignment. Fig. 9(b) shows the distance-dependent band alignment, which undoubtedly is attributed to the rutile-anatase interface interaction. Herein, our work suggests that the debate on the band alignment between rutile and anatase $TiO_2$ may be settled by a new classification of the band alignment. One class is termed the coupled band alignment containing the effect of the interface interaction between the two aligned materials, and the other is termed the intrinsic one free of that interaction.

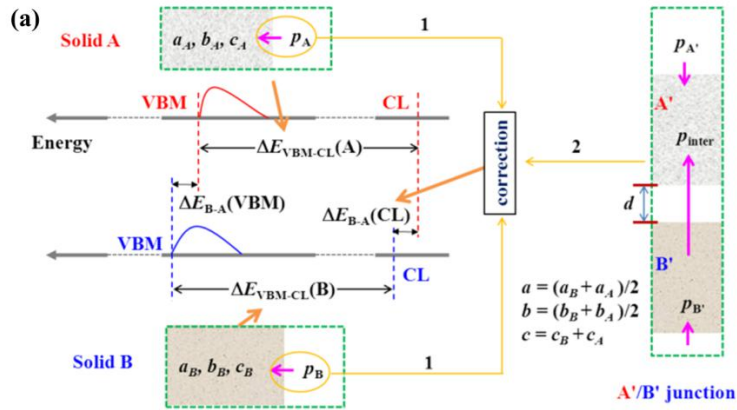

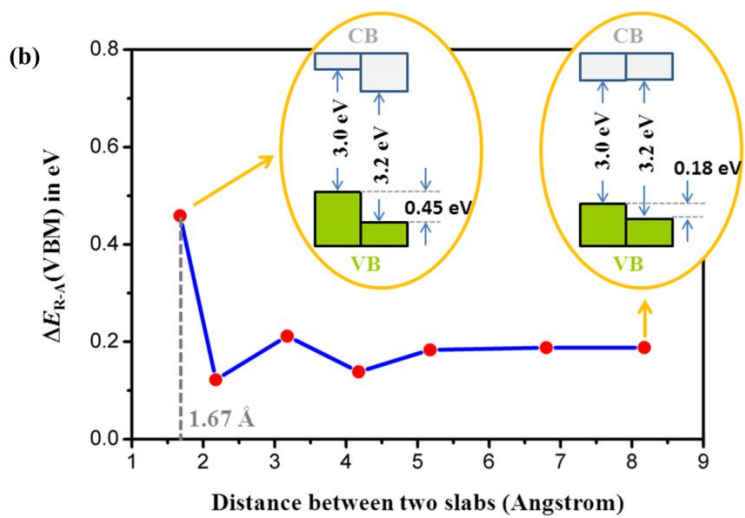

**Fig. 9.** (a) Schematic diagram of the procedure for calculating the band alignment between solids A and B. (b) The band alignment between anatase and rutile TiO$_2$ varies as the distance between two slabs. Reprinted with permission from us.[37]. Copyright © (2016) by the American Chemical Society.

**Table 1.** A collection of data on the band alignment of rutile and anatase TiO$_2$ from recent theoretical calculations and experimental observations according to a new band alignment classification proposed by us [37].

| Band alignment | Ref. | Reference energy level | Type | $\Delta E_v^*$ (eV) | Theory | Experiment |
| --- | --- | --- | --- | --- | --- | --- |
| | [24]** | flat band potential | IV | ~0 | | yes |
| | [34] | Ti 1s and O 1s core level | IV | ~0 | yes | |
| Intrinsic | [17] | charge neutrality level*** | II | 0.55 | yes | |
| | [37] | Ti 2s core level | III | 0.19 | yes | |
| | [43] | vacuum level | III | 0.15 | yes | |
| | [18] | Ti 2$p_{3/2}$ core level | II | 0.39 | | yes |
| | [39] | averaged electrostatic potential | II | 0.52 | yes | |
| Coupled | [25] | Ti 2$p$ core level | II | 0.7 | | yes |
| | [26] | work function | V | 0.2 | | yes |
| | [47] | work function | II | 0.4 | | yes |

*$\Delta E_v = E_{VBM}^{rutile} - E_{VBM}^{anatase}$ with respect to the vacuum level.

**The effect of the semiconductor/electrolyte interface on determination of the flat band potential has been cancelled using the Nernstian relation, so it is classified into the intrinsic one.

***The charge neutrality level is less rigorous for acting as the common zero energy reference.

## 5. Discussion and prospect

The band alignment between rutile and anatase $TiO_2$ has been a long-standing debate. A recent Scanlon et al.'s work on the band alignment between rutile and anatase measured by the XPS method [18], which result is opposite to the Kavan et al.'s result measured by the EIA method [24], stimulates many works on this topic. Which type of the band alignment is right? In fact, each of five band-alignment types is supported by theoretical and experimental evidences. In this sense, it seems more reasonable to reveal what main factors influence the band alignment rather than to answer which band alignment is right. We think any factor being capable of creating the macroscopic electric field will influence the band alignment. There are many such factors as defects, dangling bonds, chemisorbed functional groups at the surface, surface relaxation and reconstruction, spontaneously ferroelectric polarization, the interface of the hetero-junction, lattice deformation etc. [13, 48-54].

The formation of an interface is one of the most important factors that influence the band structures of semiconductors who compose the hetero-junction. From the knowledge of semiconductor physics, the built-in electric field created in the space charge region, similar to a macroscopic electric dipole, can lift and depress the whole band structures in the left and right sides of the junction respectively. In this way, the band alignment between rutile and anatase falls under the influence of the rutile/anatase interface. To examine this interface effect, we collect the band alignments between rutile and anatase $TiO_2$ reported by recent theoretical calculations and experimental observations, and list them into Table 1 according to whether the presence of the interface or not [37]. It follows from Table 1 that one can infer: first, the presence of an interface indeed plays a critical role in the effect on the band alignment between rutile and anatase $TiO_2$, leading to different band-alignment types shown in Fig. 2. Second, the coupled band alignment involves type II and V band alignment of Fig. 2, and so it is not strange that electron paramagnetic resonance experiments on the mixed rutile/anatase samples observed electrons flow from rutile into anatase [20, 21]. Third, the intrinsic band alignment contains type III and IV band alignment of Fig. 2, indicating the conduction band minimum energy of anatase is higher than that of rutile, and so it is natural that single anatase materials are more photocatalytic active than single rutile materials. Fourth, in that the Fermi level is a constant throughout the junction system, defects in the semiconductors will influence the band alignment as well as factors that may change the Fermi level, and so the coupled band alignments have been calculated and/or observed with a limited range of fluctuation, on which also may be the effect of other factors mentioned above or of the well-known DFT error.

Band alignment between semiconducting materials is a crucial issue in condensed matter physics and a dominant factor in renewable energy applications and electronic devices. Experiment and theory have paid extremely attention to the band alignment between applicable materials, but it still has a long way to go with challenges. Numerous factors that influence band alignment mentioned above have been informed only with a few fragmentary reports. There needs a systematic study for the effect of these factors on the band alignment. The band alignment in the presence of an interface is a more complex problem because of the dependence on the atomic structure of the interface. There were hardly any works to touch upon effects of interfacial dangling bonds, interface reconstruction, lattice mismatch etc. on the coupled band alignment. It seems to clearly understand effects of an interface on band alignment there is an urgent need of detailed information about the interface. Although this topical review only focuses on a particular case of the band alignment between rutile and anatase $TiO_2$, discussion in this paper should meet the band alignments between other semiconductors.

**Acknowledgment**

This work was supported by the National Natural Science Foundation of China (Grant Nos 11834002 and 11834002).